\documentclass[graybox,natbib,nosecnum]{svmult}
\bibpunct{(}{)}{;}{a}{}{,} 

\pdfoutput=1   

\usepackage{mathptmx}       
\usepackage{helvet}         
\usepackage{courier}        
\usepackage{type1cm}        

\usepackage{makeidx}         
\usepackage{graphicx}        
\usepackage{multicol}        
\usepackage[bottom]{footmisc}
\usepackage[normalem]{ulem}	
\usepackage{hyperref}  

\usepackage{soul}   

\newcommand{\ltsimeq}{\raisebox{-0.6ex}{$\,\stackrel
        {\raisebox{-.2ex}{$\textstyle <$}}{\sim}\,$}}
\newcommand{\gtsimeq}{\raisebox{-0.6ex}{$\,\stackrel
        {\raisebox{-.2ex}{$\textstyle >$}}{\sim}\,$}}

\newcommand{\Msun}{\mbox{M$_{\odot}$}}

\makeindex             

\begin{document}

\title*{Large-Scale Searches for Brown Dwarfs and Free-Floating Planets}
\author{Ben Burningham}
\institute{Ben Burningham \at Centre for Astrophysics Research, School of Physics, Astronomy and Mathematics, University of Hertfordshire, Hatfield AL10 9AB \email{b.burningham@herts.ac.uk}}
%
%
\maketitle

\abstract{Searches of large-scale surveys have resulted in the discovery of over 1000 brown dwarfs in the Solar neighbourhood. In this chapter we review the progress in finding brown dwarfs in large datasets, highlighting the key science goals, and summarising the surveys that have contributed most significantly to the current sample.}

\section{Introduction}
The first two confirmed brown dwarf discoveries were published in 1995 following many years of intense searching \citep{nakajima1995,rebolo1995}.   One of these, G229B, was found via high-contrast imaging as a faint companion to a nearby M dwarf \citep{nakajima1995}. The other, Tiede~1, was found from a deep imaging search (with follow-up spectroscopy) of the Pleiades cluster \citep{rebolo1995}. While both these discoveries represent the result of large efforts to identify brown dwarfs, their discovery routes turned out to be relatively minor contributors to the growing catalogue of brown dwarfs. In the following 20 years, the main discovery route for brown dwarfs was via large-scale surveys. That the first discoveries did not arise from large-scale sky surveys reflects the faintness of the targets in comparison to the depths photographic surveys available in the early 1990s, and the difficulty of identifying targets with hitherto unknown photometric properties from large catalogues. However, despite the challenge of searching for such objects in large-scale surveys, it remains the fundamental pathway for characterising the substellar population.

The first wide field searches for brown dwarfs used the all-sky photographic surveys carried out during the second half of the 20th century. 
These searches were limited by the available photometric bands (typically $BRI$), and initial ignorance of the spectral energy distributions of brown dwarfs in the solar neighbourhood. 
In addition, the depth of these surveys was insufficient to detect more than a handful of the nearest targets. 
Large-scale discovery of brown dwarfs would need to wait for the large-scale digital surveys with sensitivity in the near-infrared that came online at the turn of the 21st century. 
The discovery methods largely followed those applied to finding late-type M~dwarfs in photographic surveys, which fell into two categories: searches that relied on the motion of nearby and faint objects; and searches that distinguish the targets via their photometric colours. 
Of these, the dominant search route for brown dwarfs from wide field surveys has been photometric selection. The efficacy of the method depends on the difference between the spectral morphology of the targets of interest and the background population.
This allows the definition of a region of colour space that effectively isolates the target population. In practice, the separation is rarely perfect, and photometric selections are typically contaminated with one or more type of interloper. As a result, spectroscopic confirmation is usually required to define reliable samples.

Searches for brown dwarfs in large-scale surveys followed on from the ongoing process of extending the stellar spectral sequence to ever lower temperatures and later spectral types.  Spectral types of M7 and later are collectively known as ultracool dwarfs (UCDs). As such, all but the youngest brown dwarfs are also UCDs. The search for brown dwarfs in large-scale surveys is thus also the search for UCDs, and it both follows and drives the definition the UCD spectral sequence. Rapid progress was made extending the UCD sequence in the late 1990s and early 2000s, resulting in the definition of the L and T spectral classes \citep{kirkpatrick1999,burgasser2006}. A detailed review of these spectral classes is beyond the scope of this chapter, but may be found in \citet{kirkpatrick2005}. 
Extending the spectral sequence beyond the T sequence proved impossible using ground based surveys, and will be discussed later in this chapter, and in detail elsewhere in this Handbook.

\begin{figure}
\includegraphics[scale=0.5]{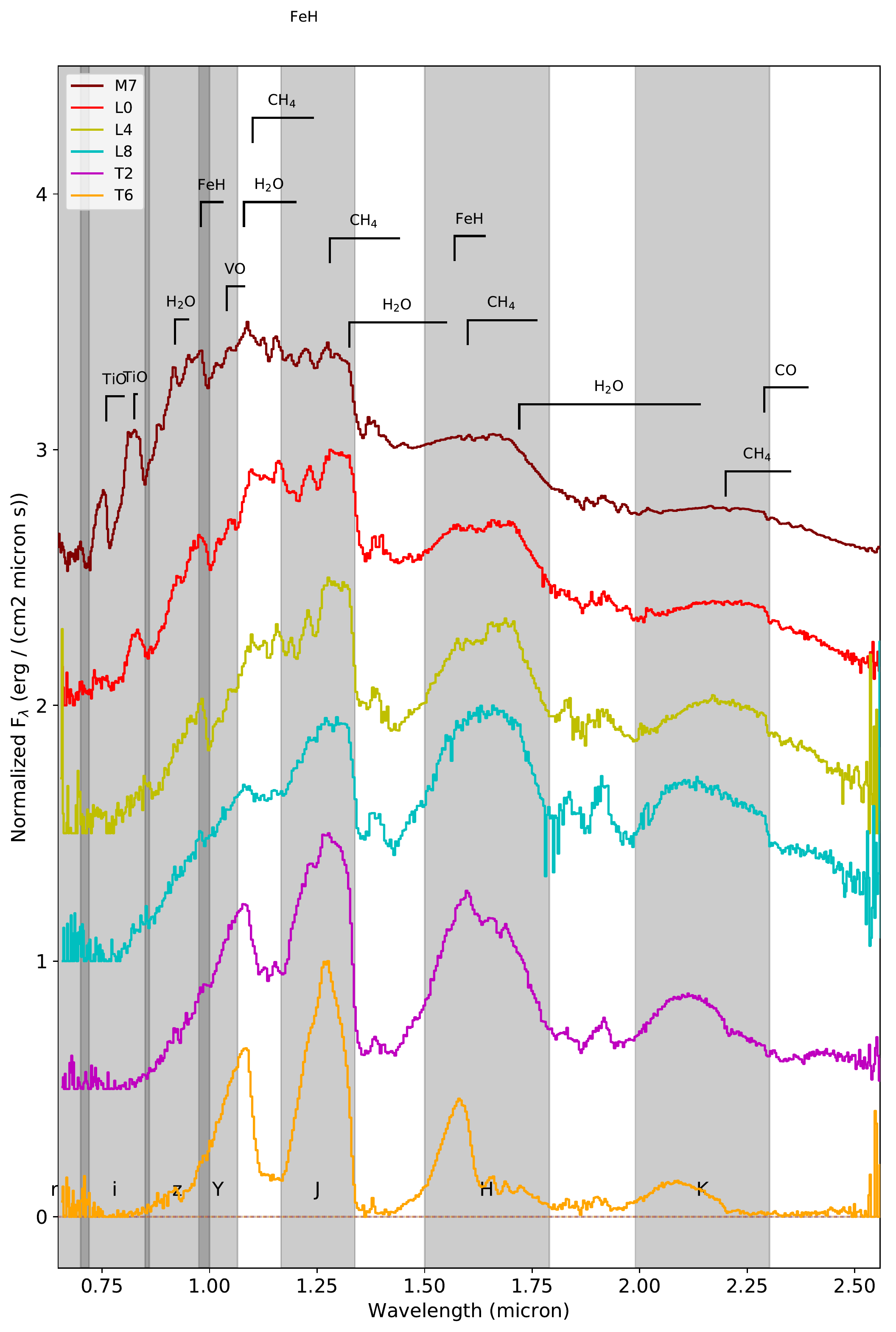}
\caption{Near-infrared spectra of M7 - T6 spectral standards from the SpeX Prism Library \citep{splat}. Key absorption features are indicated, and the approximate bandpasses for commonly used photometric filters are shaded grey.  
}
\label{fig:MLT}       
\end{figure}

Figure~\ref{fig:MLT} shows the spectral sequence from M7 to T6. The basis for key colour selection criteria is immediately apparent. 
Both L and T dwarfs may be distinguished from M~dwarfs via the red slope of their spectra as we move from optical to the near-infrared J~ band. Colour cuts such as $i' - z' \gtsimeq 1.8$ colours, or a  $z' - J \gtsimeq 2.5$ cut are often used as initial selectors for LT dwarfs.  
Often such selections require the combination of optical with near-infrared surveys, which may or may not have complementary depths. 
The faintness of the targets in the submicron region means that optical surveys often probe significantly smaller volumes for LT dwarfs than their near-infrared contemporaries. 
In these cases, the full depth of a near-infrared survey can only be searched by allowing non-detections in the optical survey to place limits in the optical to near-infrared colours.  Such search methods are known as "dropout" methods: candidates are required to be undetected at certain wavelengths.  

Typical contaminants in such selections are late-M dwarfs that have been scattered into the selection by photometric error. Such contaminants can be weeded out of searches for T dwarfs via further selection based on blue $J-H$ colours. Alternatively, selecting mid- to late-L dwarfs is facilitated by requiring very red $J-H$ colours. Selections targeting LT transition objects and early-T dwarfs suffer the greatest contamination due to $J-H \sim 0$ colours that overlap with those of M~dwarfs. As a result they have been generally underrepresented in the  samples selected from near-infrared surveys.

The nature of the photometric selection methods for LT dwarfs means that, with very few exceptions \citep[e.g.][]{folkes2012}, searches for these objects have been focused away from the Galactic plane. This is for two main reasons.  Firstly, matching survey catalogues of different wavelengths and epochs is extremely problematic in the crowded fields when the targets of interest often have large proper motions. Secondly, selecting targets based on red optical to near-infrared colour is prone to significant contamination from reddened background stars. As a result, sight lines near the Galactic plane represent the greatest source of incompleteness in the census of UCDs in the Solar neighbourhood.

Most of the dedicated large-scale searches for brown dwarfs have been led by teams within, or closely associated with, survey science teams. The details of the survey design, which may or may not have been devised with brown dwarf science in mind, drive the search strategies that these teams employ. While discovery science has undoubtedly been pursued by scientists beyond the survey consortia, the realities of winning telescope time for extensive follow-up mean that the head start given to the survey teams has often been decisive. 
Consequently, discoveries of brown dwarfs are commonly tied to one particular survey, even though multiple survey datasets are typically used to select candidates from the many millions of detected sources. 

Table~\ref{tab:totals} summarises the tally of L and T dwarf discoveries from wide field surveys over the past two decades, along with an estimate total number of {\it detectable} LT dwarfs if the full survey depth and area was exploited in each case. Each of these surveys is discussed in detail later in this chapter. 
What is immediately apparent is that the numbers of spectroscopically confirmed LT dwarfs fall well short of the maximum detectable number. This is for several reasons. 

Firstly, even low-resolution spectroscopic confirmation of LT dwarfs requires significant resources in the form 4m-class telescope time for targets with $J \ltsimeq 17.0$, or 8m-class telescope time for fainter targets. This means that the number of confirmed LT dwarfs in each survey often depends on how its volume is accessed: via wider coverage or deeper imaging over smaller areas. For example, the UKIDSS Large Area Survey (LAS) has confirmed fewer L~dwarfs than 2MASS, despite probing nearly ten times the volume. 
This reflects the fact that the extra volume probed by the UKIDSS LAS was at much fainter magnitudes. As a result, spectroscopic follow-up was largely restricted to 8m class facilities, and focused on tightly defined science goals such as understanding the historic birthrate of brown dwarfs \citep[e.g.][]{dayjones2013,marocco2015} or constraining the substellar initial mass function \citep[e.g][]{ben2013}.

The changing nature of the science that drives the search for brown dwarfs in large surveys is another big factor. 
Current community effort appears to be focused on exploring the extremes of brown dwarf parameter space such as low temperatures \citep[e.g.][]{luhman2014a, skemer2016b}, young ages and planetary masses\citep[e.g.][]{gagne2017,faherty2013}, and low metallicity \citep[e.g.][]{lodieu10,mace2013b}. 
Studying such extremes typically involves detailed follow-up of particularly interesting targets. 
The sheer expense of detailed characterisation of very faint brown dwarfs precludes their study unless faint targets are only the examples of their kind, as in the case of the coolest brown dwarfs. This limits the exploitation of deeper surveys, so all-sky surveys continue to provide the dominant resource for ongoing studies of brown dwarfs in the Solar neighbourhood.

\begin{table*}
\begin{tabular}{l c c c c c c c c}
\hline
Survey & Bands & Area / deg$^2$ & Depth & $N_{pub}$(L) & $N_{pub}$(T) & Refs & $N_{det}$(L) & $N_{det}$(T) \\
\hline
DENIS & $iJH$ & 20000 & $J = 16.5$ &49 & 1 & 1--7 & 1400 & 56 \\
2MASS & $JHK_s$ & all-sky & $J = 16.5$ &403 & 57 & 8 -- 39 & 2800 & 110  \\
SDSS (I \& II) & $ugriz$ &  15000 & $z' = 20.5$ &381 & 55 & 40 -- 50 & & \\
UKIDSS-LAS & $YJHK$ & 3600 &  $J = 19.6$ & 142 & 263 & 50--59 & 22000 & 1100 \\
CFDBS(IR) & $iz(J)$ & 1000 (355) & $z'_{AB} = 24.0; J = 20.0$ & 170 & 45 & 60--64  & 3800 & 180 \\
WISE & $W1W2W3W4$ & all-sky &  $W2 = 15.6$ &  10 & 176 & 64--67  & 19000 & 1200 \\
\hline
\end{tabular}
\caption{The numbers of published L and T dwarfs by survey, discovery references and the potential numbers detectable based on the surveys' depths and current estimates of the L and T dwarf space densities. Space densities were compiled from data in \citet{cruz2007}, \citet{dayjones2013} and \citet{kirkpatrick2012}. SDSS potential yields are not projected due to poor availability of mean $z'$ magnitudes for LT dwarfs. The CFBDS(IR) yields are based on the region with $J$ band overlap.  References: \small{ 1) \citet{delfosse1997}; 2) \citet{martin1999}; 3)  \citet{martin2010}; 4) \citet{bouy2003}; 5)  \citet{kendall2004}; 6) \citet{phanbao2008}; 7) \citet{artigua2010} ; 8) \citet{kirkpatrick1999} ;9)\citet{kirkpatrick2000}; 10) \citet{kirkpatrick2008}; 11) \citet{kirkpatrick2010}; 12) \citet{burgasser1999} ; 13) \citet{burgasser2000};14) \citet{burgasser2002}; 15) \citet{burgasser2003} ; 16) \citet{burgasser2003a} ; 17) \citet{burgasser2003b}; 18) \citet{burgasser2004a};19) \citet{burgasser2004b} ; 20) \citet{reid2000} ; 21) \citet{reid2008}; 22)  \citet{gizis2002a}; 23) \citet{gizis2000}; 24) \citet{gizis2003}; 25) \citet{kendall2003}; 26) \citet{kendall2007}; 28) \citet{cruz2003}; 29) \citet{cruz2004}; 30) \citet{cruz2007}; 31) \citet{wilson2003} ; 32) \citet{folkes2007}; 33) \citet{metchev2008}; 34) \citet{looper2007}; 35) \citet{looper2008};  36) \citet{sheppard2009}; 37) \citet{scholz2009}; 38) \citet{geissler2011}; 39) \citet{tinney2005}; 40) \citet{fan2000}; 41) \citet{hawley2002}; 42)  \citet{geballe2002}; 43)  \citet{schneider2002}; 44)  \citet{knapp2004}; 45)  \citet{chiu2006}; 46)   \citet{zhang2009}; 47)  \citet{scholz2009}; 48) \citet{schmidt2010}; 49) \citet{leggett2000}; 50) \citet{lodieu2007}; 51) \citet{pinfield2008}; 52) \citet{ben2008}; 53) \citet{ben2009}; 54) \citet{ben2010a}; 55) \citet{ben2010b}; 56) \citet{ben2013}; 57) \citet{cardoso2015}; 58) \citet{dayjones2013}; 59) \citet{marocco2015}; 60) \citet{cfbds} 61) \citet{reyle2010}; 62) \citet{cfbdsir}; 63) \citet{loic2011}; 64) \citet{kirkpatrick2011}; 65) \citet{kirkpatrick2012}; 66) \citet{mace2013}; 67) \citet{pinfield2014}; 68) \citet{lodieu2012b}}
\label{tab:totals}
}
\end{table*}

Although significant questions remain regarding the properties of the brown dwarf population, particularly in a Galactic context, there is currently little appetite within the community for addressing these questions via large-scale spectroscopic follow-up of faint brown dwarfs selected from current or future surveys. 
The work of \citet{skrzypek2015,skrzypek2016} is worth noting here. They combined eight filter bandpasses ($izYJHKW1W2$) across three surveys (SDSS, UKIDSS and WISE) to estimate photometric spectral types for some 1361 LT dwarfs. Limited spectroscopic follow-up suggests that their method is sound, and their photo-typing method achieves $\pm 1$ subtype accuracy across the LT range. Such approaches provide the opportunity to extend current studies of brown dwarfs to much larger samples and likely represent a key future methodology for large-scale searches for brown dwarfs. 

In the remainder of this chapter, we summarise the details of the surveys which contributed most significantly to the current brown dwarf census, and highlight some of the key science results of large-scale searches for brown dwarfs.

\section{Significant large area surveys}
\label{sec:las}

\subsection{The Deep Near-Infrared Survey of the Southern Sky (DENIS)}

Carried out between 1996 and 2001 the 1m-ESO telescope at La Silla (Chile),  the Deep Near Infrared Survey of the Southern Sky (DENIS) was one of the first substantial surveys in the near-infrared. It covered 20,000 square degrees in $I$ ($\lambda \approx 0.8 \mu$m), $J$ ($\lambda \approx 1.25 \mu$m) and $K_s \approx 2.15 \mu$m) photometric bandpasses \citep{denis}, and was a significant contributor to the early progress of brown dwarf science the Solar neighbourhood.  Its choice of filters provided leverage on the extremely red $0.8 - 1.2 \mu$m SED of L and T dwarfs, and so brown dwarfs could be directly selected from the survey catalogue without the need to cross-reference other surveys sensitive to other wavelengths.  This resulted in the discovery of 50 brown dwarfs (see Table~\ref{tab:totals}).

\subsection{The Two Micron All Sky Survey (2MASS)}

The Two Micron All-Sky Survey \citep[2MASS ;][]{2mass} was the first, and to date only, all-sky survey covering the $1- 2.5 \mu$m near-infrared region. A transformational contribution to the study of low-mass stars and brown dwarfs, it provided discovery images for over 400 L~dwarfs and nearly 60 T~dwarfs (Table~\ref{tab:totals}) and continues to feature as a key dataset in many ongoing studies of brown dwarfs in the Solar neighbourhood. The survey was completed between 1997 and 2001 using one 1.3m telescope in each hemisphere: one at the Fred Lawrence Whipple Observatory, on Mount Hopkins, Arizona (USA); and one at the Cerro Tololo Inter-American Observatory (Chile).  The survey imaged the whole sky in three filters: $J$~($\approx 1.235 \mu$m); $H$~($\approx 1.662 \mu$m); $K_s$~($\approx 2.159 \mu$m).  


Alone, the wavebands covered by 2MASS would not allow photometric selection of brown dwarfs against a background of stars with similar $JHK_{s}$ colours. However, its depth in $JHK_{s}$ was well matched to the depth of the various photographic surveys that were digitised during the 1990s, allowing the selection of candidate brown dwarfs using the dropout technique. The principal search strategy for brown dwarfs in 2MASS was photometric and relied on the red optical-to-NIR colours of L and T dwarfs, combined with their respective red and blue $J - K_{s}$ colours. Initial candidate selections required non-detections in the red photographic plates,which combined with 2MASS detection limits to set a colour limit of $R - K_{s} \gtsimeq 5.5$ \citep[e.g.][]{kirkpatrick1999}. The NIR colours of candidates were then used to distinguish T~dwarfs and L~dwarfs. 
The many discoveries made in 2MASS are summarised and referenced in Table~\ref{tab:totals}.



\subsection{The Sloan Digital Sky Survey (SDSS)}
The SDSS is widely regarded as being the most successful astronomical survey of all time. The SDSS really represents a set of surveys that continue through the time of printing, targeting diverse science goals. Here, we consider the original SDSS obtained as part of SDSS I \& II, now known as the SDSS Legacy Survey. Data collection started in 2000 \citep{sdss} and the final data release took place in 2011 \citep{sdssdr8}, incorporating data taken as part of the original SDSS observing plan up until 2008.   The facility has diversified its survey portfolio since 2008, pursuing spectroscopic surveys targeting wide-ranging science goals including exoplanets, Galactic structure and the large-scale structure of the Universe across optical to near-infrared wavelengths \citep[e.g.][]{sdss3,sdss4}.  The SDSS Legacy Survey consisted of two principal components: a photometric survey and a spectroscopic survey. Both were obtained using the 2.5-m wide-angle optical telescope at Apache Point Observatory in New Mexico (USA).

The SDSS photometric survey imaged some 8,000 square degrees of sky in $ugriz$ filters. The survey region targeted the northern Galactic cap, and three stripes in covering the southern galactic cap. This strategy minimised contamination from Milky Way foreground gas, dust and stars that would interfere with the survey's principal goal of constructing a three-dimensional map of the distribution of galaxies. This strategy did not hinder searches for brown dwarfs in the solar neighbourhood, which would generally avoid the Galactic plane in any case.   The spectroscopic survey was predominantly targeted at determining redshifts for galaxies, and it obtained spectra of some 1.8 million targets 

Brown dwarfs were discovered within the SDSS photometric catalogues from the outset, with seven L~dwarfs and two T~dwarfs identified in commissioning data \citep{strauss1999,tsvetanov2000,fan2000}. Candidates were selected via $(i' - z')$ vs $(r'-i')$ colour-colour diagrams, and via $r'$ and $i'$ band dropout searches. Searches of SDSS photometric catalogues were also complemented with 2MASS photometry to further constrain the colours of the targets \citep[e.g. ][]{chiu2006}. SDSS provided the discovery data for some 381 L~ dwarfs and 55~T~dwarfs to-data (see Table~\ref{tab:totals}.

In addition to photometric selections of candidates, the work carried out by \citet{schmidt2010} is of note for selecting L~dwarfs based on their spectra, rather than broadband colours. This unique selection method is largely free of colour biases that can be introduced by photometric methods, and was made possible thanks to the spectroscopic survey carried out as part of the SDSS. Although the vast majority of  targets within the SDSS spectroscopic survey were extragalactic in nature, approximately 5\% of its spectra were of objects with late-M spectral type and cooler \citep{schmidt2010}.  The resulting sample of {\it spectroscopically} selected L~dwarfs had a median $J-K_{s}$ colour 0.1 magnitudes bluer than previous photometric selections for spectral types on the L0 - L4 range. This example highlights how colour based selections can introduce bias, particularly when colour cuts are aimed at distinguishing objects in spectral type transition regions.

\subsection{The UKIRT Infrared Deep Sky Survey (UKIDSS)}

The UKIRT Infrared Deep Sky Survey (UKIDSS) was carried out using the purpose built Wide-Field CAMera \citep[WFCAM; ][]{wfcam} between 2005 and 2012.   UKIDSS employed what is commonly known as a wedding cake strategy: the survey as a whole consists of a set of sub-surveys of decreasing sky-area and increasing depth \citep{ukidss}. The original plan for UKIDSS aimed to image some 7500 square degrees, but the largest planned area for single sub-survey was just over half that area. This approach is seen frequently in modern sky surveys which tension various science goals with differing requirements in terms of depth and coverage against finite observing time on dedicated survey instruments. 

In order of sky coverage, the surveys that comprise UKIDSS are as follows.  The Large Area Survey (LAS): 3700 sq. degs. in $YJHK$ principally covering overlap sky with SDSS outside the Galactic plane to a typical $5\sigma$ depth of $K < 18.4$.
The Galactic Plane Survey (GPS): 1800 sq. degs. at  $JHK$ covering the Galactic plane within $b \pm 5\deg$ to $K < 19.0$.
The Galactic Clusters Survey (GCS): 1400 sq. degs. covering 10 star clusters in $JHK$ to a depth of $K < 18.7$).  The Deep Extragalactic Survey (DXS): 35 sq. degs. in $JHK$ to $K < 21$). The Ultra Deep Survey (UDS): 0.8 sq. degs. in $JHK$ to $K < 25.3$. 


The most prolific of the UKIDSS surveys for brown dwarf detections was the LAS. The top two LAS headline science goals were: discovering the coolest brown dwarfs in the Solar neighbourhood, and identifying the highest redshift quasars ($z > 6$). Both of these target populations are heavily contaminated by Galactic M~dwarfs in NIR colour-colour diagrams, so the bulk of the LAS footprint was placed to coincide with the SDSS. This allowed the reddest bands of the SDSS to be used to exclude the populous M~dwarfs to nearly the full depth of the $J$ band survey. 

To further aid in the photometric selection of its key science targets, the LAS was the first wide field survey to employ the MKO $Y$ band filter, centred at $1.02 \mu$m. This filter was designed to allow effective discrimination between high-redshift quasars and T(+)~dwarfs, which otherwise share similar colours in $JHK$.  This discrimination relied on the fact that the L~ and T~dwarfs discovered up to this time had $Y - J > 1.0$, with cooler brown dwarfs expected to be even redder, whilst high-redshift quasars were expected to remain bluer than this limit up to $z \sim 7$ \citep{warren2002}.  Early searches of the LAS for extremely cool T dwarfs, and the preemptively classified Y~dwarfs, were guided by this expectation. However, the first brown dwarf to be identified with spectral type later than T8, ULAS~J0034-0052, was excluded from early searches and was instead identified as part of a search for quasars,  displaying $Y - J = 0.75 \pm 0.1$.  

As the survey progressed it became clear that late-T~dwarf colours frequently overlapped with those of high-redshift quasars, with the latest type objects generally found with $Y - J < 1$ \citep[e.g.][]{ben2010b,ben2013}. This trend to bluer $Y - J$ colours is now understood in terms of rain-out chemistry removing {\sc K I} from the gas phase in the coldest objects, and weakening the strong, pressure-broadened, potassium line that dominates the T dwarf spectral morphology in the $Y$ band \citep[e.g.][]{line2017}. The small number of T7+ dwarfs discovered in previous surveys did not initially show this trend, and the experience serves to illustrate the potential pitfalls of using previously justified photometric selection criteria to explore new parameter space for ultracool dwarfs. 

The final photometric selection methods for the principal UKIDSS late-T dwarf follow-up programme are outlined in detail in \citet{ben2013}, which also provides copies of the SQL queries used to the perform the selections in the WFCAM Science Archive (WSA).  These selections employed a relatively weak $Y - J > 0.5$ requirement.  This weak criterion was necessary to avoid excluding late-T dwarfs with blue $Y-J$ colours, but let many M~dwarfs pass the selection.  A $J-H < 0.1$ cut removed the bulk of L and M dwarfs, while a final $z' - J > 2.5$ excluded further M~dwarfs. For the bulk of the volume searched, this final cut was achieved by requiring candidates to the undetected in the SDSS.  This search confirmed some $\sim 200$ T dwarfs, making it the most prolific source of confirmed T~ dwarfs to-date.  

\subsection{Canada-France Brown Dwarf Survey CFBDS(IR)}

The Canada-France Brown Dwarf Survey \citep[CFBDS;][]{cfbds} covered some 1000 square degrees in $iz$ filters, with $J$ band coverage in 355 square degrees. The $iz$ survey drew data from two existing surveys carried out using MegaCam on the Canada-France-Hawaii Telescope (CFHT) on Mauna Kea, Hawaii (USA): the CFHT Legacy Survey \citep[CFHTLS;][]{cfhtls} and the Red Sequence Cluster Survey \citep[RCS-2;][]{rcs2}. 
Candidate brown dwarfs were selected on the basis of red $i - z$ colours, and followed up with $J$ band imaging. 
Despite its modest coverage this survey made a significant contribution to determining the local space density of brown dwarfs \citep{reyle2010} and discovering extremely cool T dwarfs \citep{delorme2008}.

\subsection{The Wide-field Infrared Survey Explorer (WISE)}

Launched in 2009, the WISE mission surveyed the entire sky in four wavebands that are largely inaccessible from the ground, centred at 3.4$\mu$m (W1), 4.6$\mu$m (W2), 12$\mu$m (W3) and 22$\mu$m (W4) \citep{wise}. In many ways in can be viewed as a successor to 2MASS in terms of brown dwarf science. Early exploitation and follow-up for brown dwarf science was led by the same team at the Infrared Processing and Analysis Center in Pasadena that pursued early brown dwarf science with 2MASS. As with 2MASS, early science drivers were concerned with completing a census of brown dwarfs in the solar neighbourhood and defining a new spectral type, in this case the Y spectral type. 

The initial survey design was based around a cryogenic mission lifetime of 6 months, during which the full sky was imaged in all four passbands. The cryogen lasted after the initial pass was completed, allowing 20\% of the sky to be imaged a second time in all four bands. The shortest two wavebands (W1 \& W2) do not require additional active cooling of the spacecraft by venting cryogen, and post-cryogen operations were planned to take advantage of this.  Following the exhaustion of cryogen, the mission was renamed NEOWISE (for Near-Earth Object WISE), and completed the second pass of the whole sky in W1 and W2 with the aim of detecting potentially hazardous NEOs \citep{neowise}. The point source catalogue for the full cryogenic mission was released as the WISE All-Sky data release in 2012. In 2013, data from the cryogenic mission and subsequent post-cryogenic NEOWISE mission were published as the ALLWISE data release.

The WISE spacecraft performed the survey by scanning the sky as it orbited the Earth above the terminator region, building up depth via multiple passes over each region. The regions near the ecliptic poles thus received the greatest number of images (1000s of images at the poles), whereas regions on the ecliptic place typically received 12 to 13 passes in the original cryogenic mission.  Moon avoidance led to some regions receiving considerably fewer passes.

The WISE all-sky $W1$ and $W2$ depths probe a similar volume for L~dwarfs to that probed by 2MASS, however as one moves to cooler temperatures the probed volume soon overtakes that probed by 2MASS for T dwarfs.  As a result, the brown dwarf discoveries by WISE are dominated by objects with late-T type and beyond \citep[e.g. ][]{kirkpatrick2011, mace2013}. 
Although its probed volume for late-T dwarfs is similar to that of the UKIDSS LAS, WISE provides a more accessible sample due to its all-sky coverage giving a greater volume at smaller distances. Nonetheless, the comparable discovery number of T dwarfs in UKIDSS LAS and WISE largely reflects shifting science priorities, and the lack of clear motivation for spectroscopic follow-up of large numbers of faint mid- to -late-T dwarfs.   
The early searches of the WISE dataset were also successful in discovering the long sought Y~dwarfs \citep{cushing2011}.

The presence of multiple passes and two full sky surveys notably facilitated NEO detection, but also allows for efficient searches for high proper motion objects beyond the Solar System. 
Of particular note are the efforts of Kevin Luhman at the Pennsylvania State University Center for Exoplanets and Habitable Worlds, whose proper motion searches using WISE's multi-epoch imaging has resulted in the discovery of some of the Sun's closest substellar neighbours \citep{luhman2013, luhman2014a}.  An L7.5+ T0.5 binary \citep{burgasser2013} at a distance of less than 2~pc \citep{sahlmann2015}, Luhman 16AB was visible in previous surveys DSS, 2MASS and DENIS but was not identified as a nearby brown dwarf system due to confusion with other nearby sources. By contrast, WISE~J085510.83-071442.5, at a distance of 2.2~pc \citep{luhman2014a,luhman2016} lacked a ground-based detection for nearly two years following its discovery \citep[e.g.][]{skemer2016b}. This reflects the tiny flux emitted at wavelengths easily accessible from the ground by this extremely cool brown dwarf, with an estimated $T_{\rm eff} \approx 200 - 250$K \citep{schneider2016}.

Whereas Luhman's strategy relied on fairly cumbersome searching the single-exposure catalogues, followed by labour-intensive vetting of candidates by eye,  the ALLWISE catalogue provides a more convenient route to leveraging the WISE extended dataset to search for high-motion objects. The ALLWISE catalogue forms the basis for a number of motion-based surveys for brown dwarfs in the solar neighbourhood \citep[e.g.][]{kirkpatrick2014}.

In December 2013, the WISE spacecraft was reactivated and NEOWISE survey operations continued through Spring 2018.  The increasing depth and motion data raise the likelihood of further exciting discoveries from the WISE spacecraft. Ongoing efforts to mine these data for brown dwarfs include the citizen science project "Backyard Worlds", which has found at least one cool brown dwarf to date \citep{kuchner2017}.

\subsection{The Visible and Infrared Survey Telescope for Astronomy (VISTA)}
The European Southern Observatory's (ESO) Visible and Infrared Survey Telescope for Astronomy \citep{vircam} has facilitated a number of public surveys, the first tranche of which follow a similar wedding cake design to that seen for the UKIDSS surveys. 
Three of the public surveys hold particular potential for brown dwarf science: the VISTA Hemisphere Survey \citep[VHS; ][]{vhs}; the VISTA Kilo-degree survey \citep[VIKING;][]{viking}; the VISTA Variables in the Via Lactea  \citep[VVV; ][]{vvv}.

The first of these has delivered a handful of brown dwarfs found following similar photometric methods to those applied to exploitation of UKIDSS \citep[e.g][]{lodieu2012b}. 
One of the reddest known L~dwarfs, VHS~J1256601.92-125723.9, was identified in the VHS as a companion to a brown dwarf binary with a probable age of 300~Myr \citep{gauza2015,stone2016}.
However, although the nearly half-sky coverage of VHS in the near-infrared provides excellent opportunities for large-scale searches for brown dwarfs, it initially lacked the complimentary optical coverage that is so important for selecting LT dwarfs. 
The 1500 square degree $ZYJHK$ VIKING survey provides opportunity to select LT dwarfs in a self-sufficient way, but the limited coverage means that the candidates will be faint, and require expensive follow-up.

The VVV has provided one of the best opportunities to date to search within the Galactic plane for nearby brown dwarfs. 
This survey covers the southern Galactic plane in $ZYJHK_s$, but with around 100 epochs in the $K_s$ filter over a 7-year period \citep{vvv}. 
This multi-epoch survey has been optimised for studying variable stars, however, it also allows astrometric selection of fast-moving and nearby brown dwarfs that are otherwise hard to spot in the crowded Galactic plane \citep[e.g.][]{beamin2013}.

\subsection{The Panoramic Survey Telescope and Rapid Response System (Pan-STARRS)}

The Panoramic Survey Telescope and Rapid Response System \citep[Pan-STARRS; ][]{kaiser2010} is a wide-field imaging facility based around a 1.8 metre telescope on Haleakala, Maui (USA). 
 The design of the facility is based around rapid scanning of the sky to allow identification of fast-moving sources, transients and solar system objects. 
 Several synoptic surveys have been carried out with Pan-STARRS, but the survey of most interest for brown dwarf searches is the Pan-STARRS1 3$\pi$ survey \citep[PS1: ][]{ps1}. 
 Named for its coverage of the three quarters of the sky that it can access, it incorporates around 12 visits in each of its five filters ($grizy$) over the four years it took to complete (2010 - 2014).  It's typical $5\sigma$ detection threshold is $z_{ps1} \approx 22.3$.

 The optical coverage of PS1 has been combined with WISE data to target LT transition objects that have been otherwise difficult to identify in near-infrared surveys, and this is one the most significant contributions that PS1 has made to the census of local LT dwarfs \citep{best2015,best2018}. 
 The multiple epochs also provide accurate proper motions, and many brown dwarfs have also been discovered as wide common proper motion companions to stars using PS1 \citep[e.g.][]{deacon2014,deacon2017}.
 PS1 is also noteworthy for the discovery of the planetary mass brown dwarf PSO~J318.5-22  \citep{liu2013}.

\section{Science drivers for large-scale brown dwarf searches} 
\label{sec:science}
The science drivers behind large-scale searches for brown dwarfs have evolved over the years, from identifying first examples of new classes of objects, to growing statistically useful samples for population studies, targeting outliers and expanding parameter space.  
Here we explore some of the key science goals that have driven searches for brown dwarfs via large-scale surveys in the past few decades.

\subsection{The Initial Mass Function}

Much of the popular excitement surrounding the search for brown dwarfs in the 1990s was thanks to idea that brown dwarfs might account for a significant component of dark matter, under the umbrella of Massive Compact Halo Objects (MACHOs). 
However, there were already good reasons to suspect that brown dwarfs at most accounted for a small proportion of dark matter. 
This did not significantly reduce the impetus for confirming the existence of the hitherto unseen population, since (for many) the real motivation for this search was an understanding of the star formation process via a full accounting of its products.  

The initial mass function \citep[IMF;][]{salpeter55} describes the rate of star formation as a function of mass, and is often thought of as the distribution of mass in a coeval stellar population as a function of stellar mass. It has long been regarded as one of the fundamental diagnostics for testing models of star formation, despite the impossibility of measuring it directly over the full stellar and substellar mass range. This measurement is challenging for many reasons.  

Prime of these is the fact masses are not directly observed, so the IMF is inferred from the luminosity function (LF).
Although the mass-luminosity relationship is generally well characterised for main sequence stars in hydrostatic equilibrium,
it is not possible to observe coeval populations of main sequence stars over the full mass range: the most massive stars have exploded as supernovae before the lowest-mass stars have reached the main sequence.  
Determining masses for pre-main sequence (PMS) stars in young coeval populations depends on PMS evolutionary models and correctly determining the age of the population. Both of these hurdles introduce significant uncertainty to the resulting IMF.

These issues can be avoided to by considering the field population, as Salpeter did in his seminal paper that first defined the IMF \citep{salpeter55}. Instead, one must account for rate of stars evolving off the main sequence, and the star formation history of the field population.  In the case of the substellar population, which never reaches the main sequence and thus lacks a unique mass-luminosity relationship in a mixed age population, reference must also still be made to evolutionary models to determine object masses. 
Salpeter was untroubled by this last point: his luminosity function corresponded to a mass range of roughly $0.4 - 10 M_\odot$. His mass distribution was well fit by a power law, $\zeta (m) \propto m^{-\alpha}$, with $\alpha = 2.35$, which is now known as the Salpeter mass function. The human story behind this first derivation of the IMF is wonderfully described in \citet{salpeter2005}.  

The essential quality of the Salpeter mass distribution is that the number of stars increases steeply with decreasing mass. Moreover, since $\alpha > 2.0$, it reflects more mass being sequestered in lower-mass stars than higher-mass stars when integrated over equal logarithmic mass bins. 
Extrapolating this to ever lower masses leads to the prediction that brown dwarfs (and planetary mass objects) might represent a dominant constituent of baryonic matter in the Galaxy.  
However, in the following years it became clear that the mass distribution flattened below 1\Msun, and by the 1990s it was clear that a significant upturn in the substellar regime would be required for brown dwarfs to be numerous enough to account for dark matter \citep{sandage1957,schmidt1959,miller1979,scalo1986}. 
An in-depth review of progress in constraining the stellar IMF is beyond the scope of this chapter, so the reader is directed to an excellent review by \citet{bastian2010}. Continued studies of the IMF in young clusters, globular clusters, other galaxies and the local field support the idea that the IMF is apparently universal across much of the stellar mass range, regardless of environment \citep[e.g.][]{scalo1986,bastian2010}. 
There is consensus that the Salpeter IMF holds for masses greater than about 1\Msun. 
For masses below $\sim \Msun$,  however, the IMF can be fit by shallower power law \citep[e.g. $\alpha \sim 1.0 - 1.3$][]{reid2002} or log-normal form for low-mass stars \citep{chabrier2003}. 

\begin{figure}
\includegraphics[scale=1.]{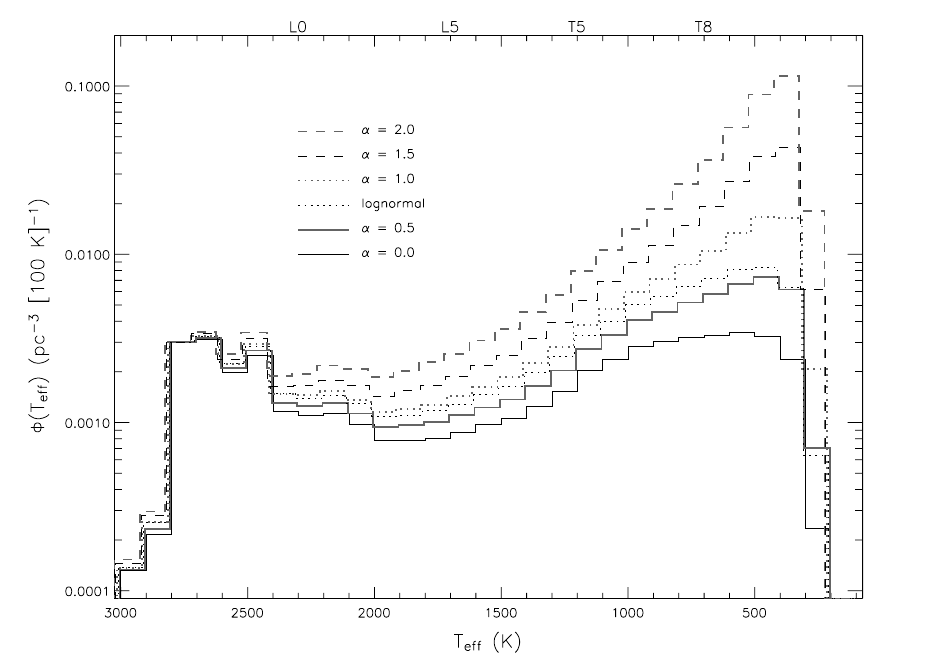}
\caption{From \citet[][Fig. 5]{burgasser2004}. Simulated $T_{\rm eff}$ distributions for the local field population under different assumed IMF forms for historically constant formation rate.}
\label{fig:IMFsim}       
\end{figure}

The first extension of the luminosity function to ultracool temperatures, and substellar masses, was facilitated by the Two Micron All-Sky Survey \citep[2MASS; see Section~\ref{sec:las};][]{2mass}, with key papers by \citet{cruz2003, cruz2007} which robustly explored the LF across the M7 - L8 spectral-type range. However, studies of this spectral-type range had limited value for constraining the IMF.
The lack of unique mass-luminosity relationship for brown dwarfs significantly complicates the estimation of the form of IMF in the mixed age field population.  Since the age, and hence the mass,  for isolated brown dwarfs cannot be reliably determined,  estimates for the IMF depend on comparisons of observed LFs or spectral-type distributions to simulations based on different assumed IMFs and historic formation rates. Figure~\ref{fig:IMFsim} shows simulated $T_{\rm eff}$ distributions under a range of assumed IMF power laws and a log-normal IMF from \citet{burgasser2004}.  The luminosity function in the late-M and early-L spectral-type range shows relatively weak dependence on the form of the underlying mass function. 
As such, the well-measured space densities in this regime placed weak constraints on the form of the substellar IMF, with \citet{allen2005} finding the LF consistent with $\alpha = 0.0 \pm 0.5$.

From Figure~\ref{fig:IMFsim} it is clear that the sub-1000~K late-T dwarf $T_{\rm eff}$ distribution carries the greatest potential for constraining the slope of the substellar IMF in the field. Although 2MASS and SDSS were responsible for defining the T dwarf spectral sequence, they lacked the depth to detect a sufficient number of T6+~dwarfs to constrain the IMF with any statistical power. 
For this reason, late-T dwarfs were preferentially targeted for large-scale searches as soon as the probed volume allowed for the selection of useful samples. The UKIDSS survey was the first to achieve this, and it was found that the space density of T6 - T8 dwarfs was most consistent with a steeply declining substellar IMF, with $\alpha < 0.0$ \citep{pinfield2008, ben2010b,ben2013}. This was subsequently confirmed in the WISE census of the Solar Neighbourhood \citep{kirkpatrick2012}.
This is in contrast to a number of determinations for the substellar IMF in young clusters and associations which have generally found $\alpha > 0.0$, e.g. Upper Sco, 0.3--0.01\Msun,  $\alpha = 0.6 \pm 0.1$ \citep{lod07a}; Pleiades, 0.48--0.03\Msun, $\alpha = 0.60 \pm 0.11$ \citep{moraux03}; $\alpha$ Per, 0.2--0.04\Msun, $\alpha = 0.59 \pm 0.05$ \citep{barrado02}; $\sigma$ Orionis, 0.5--0.01\Msun, $\alpha = 0.5 \pm 0.2$ \citep{lodieu09}; $\sigma$ Orionis, 0.25--0.004\Msun, $\alpha = 0.6 \pm 0.2$ \citep{penaramirez2012}.


The reason for the discrepancy between the substellar IMF estimated in the field and that estimated in young clusters is not clear. 
Trivial incompleteness in the field studies is an unlikely origin of the discrepancy since the surveys would need to miss more late-T dwarfs than they found to account for the difference.
Another possibility is incorrect treatment of the historic substellar formation rate when simulating the IMF, which has generally assumed a flat formation rate \citep[e.g.][]{ben2013}. For example, a low historic formation rate might give rise to an under-abundance of late-T dwarfs in the Solar neighbourhood, despite sharing the young cluster IMF.
 However, studies of the kinematics of the late-T population suggest that it is of a similar age to the stellar population on the Solar neighbourhood \citep{smith2013}. Similarly, work by \citet{dupuy2017} supports the assumption of a relatively flat formation history for brown dwarfs. 
Another possible cause is some issue with the evolutionary models used to transform between mass and temperature at young ages or over Gyr timescales. 
Alternatively, the form of the IMF may deviate significantly from a power law or log normal form below the masses probed in young clusters. In that case, simulations of the mixed age field population based on such assumed forms may be expected to disagree with observed space densities in the field due to influence from the mass population below the sensitivity of previous cluster studies.

\subsection{Extending the spectral sequence to ever lower temperatures}
One of the headline science goals for large area surveys in the period following 2MASS and SDSS was the discovery of objects cooler than the T8 ($T_{\rm eff} \approx 700$K) low temperature extent of the T spectral sequence defined in \citet{burgasser2006}.  
Speculation was divided over the question as to whether another spectral type would be required beyond the T sequence, or if the T dwarfs would be final entry in the stellar spectral classification scheme.
Comparisons of the coolest T dwarfs with Saturn and Jupiter suggested that ammonia absorption should be become increasingly important with decreasing temperature, and that new features in the $Y$ and $J$ bands might distinguish a new spectral sequence \citep[e.g.][]{leggett2007}.  
Another speculated driver for a shift in the spectral sequence was the impact of water clouds condensing at $T_{\rm eff} \ltsimeq 400 - 500$~K.
Regardless of the ultimate rationale for adopting a new scheme, the class beyond the T~dwarfs was pre-emptively named the Y~dwarfs  \citep{kirkpatrick1999}. 

Two surveys in the mid-2000s were targeted at identifying objects beyond the T~sequence: UKIDSS and CFBDS(IR). 
Both surveys were successful at identifying cooler objects than had been found previously, e.g.: CFBDS~J005910.90-011401.3 at $T_{\rm eff} \approx 620$K \citep{delorme2008}; UGPS~J072227.51-054031.2 at $T_{\rm eff} \approx 520$K \citep{lucas2010}. 
However, even at $T_{\rm eff} \approx 500$K, the objects' spectra continued to appear as a continuation of the T spectral sequence. 
Although the methane bands continued to strengthen with approximately similar relative changes between subtypes, the absolute changes were small and strongly argued for the continuation of the T~sequence for these new objects \citep[e.g.][]{ben2008,ben2010b,lucas2010}. 
More recent analysis has highlighted the measurable impact of ammonia on the near-infrared spectra on objects with spectral types T8 and later \citep[e.g.][]{line2015, canty2015}, but its effect at these temperatures does not cause a qualitative deviation from the T~sequence. 

\begin{figure}
\includegraphics[scale=.6]{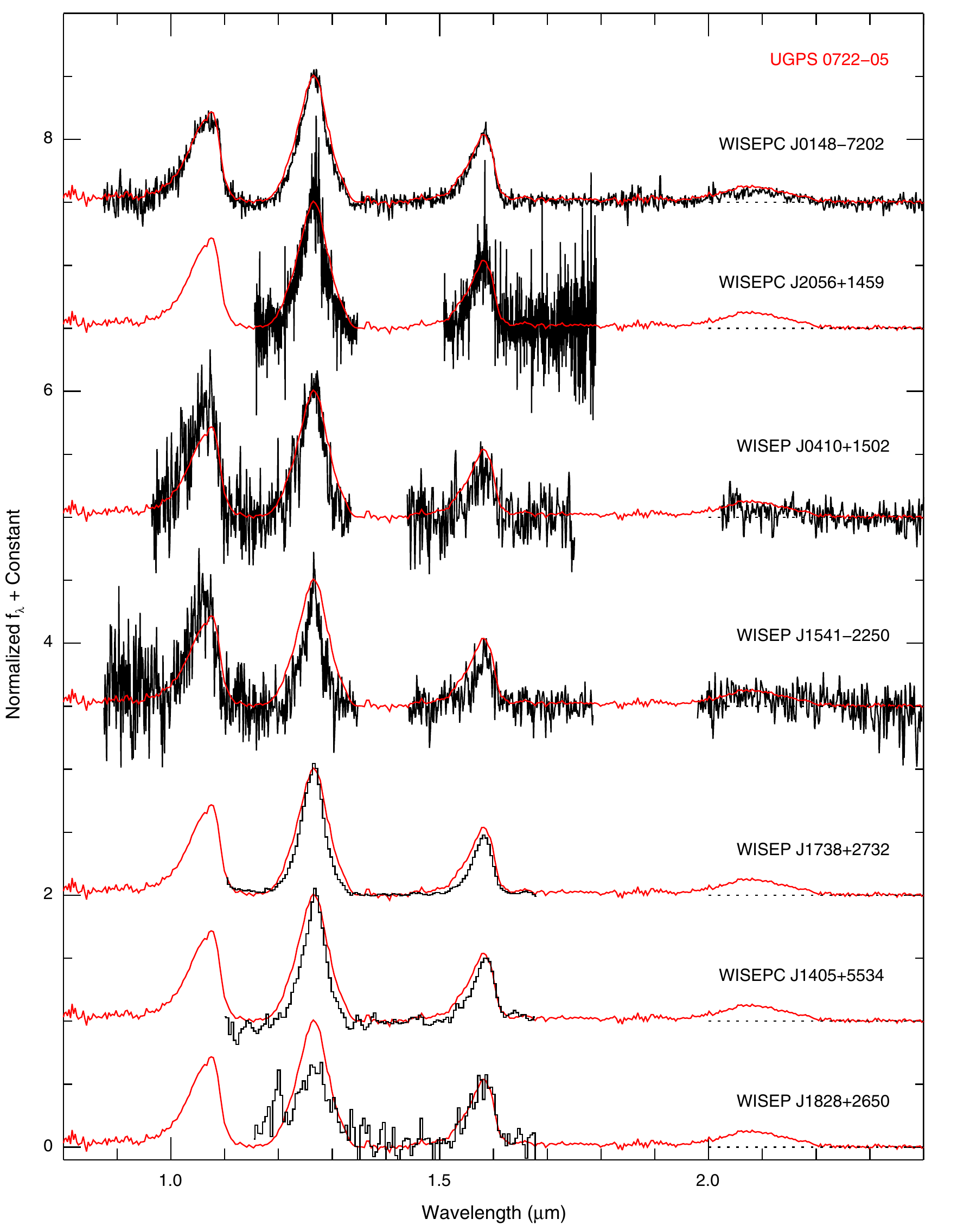}
\caption{From \citet[][Fig. 2]{cushing2011}. Spectra of the first Y dwarfs found in the WISE survey. Note the comparable heights of the flux peaks in the $Y$ and $J$ bands compared to the T9 standard UGPS~J0722-05. Also apparent is the narrower $J$ band peak in the $Y$ dwarfs.  }
\label{fig:Yspec}       
\end{figure}

The launch of the WISE spacecraft provided the necessary sensitivity to identify even cooler objects, which would justify the adoption of a new spectral type. 
Selected by their brightness at W2 above all else, the Y~dwarfs are differentiated from the T~sequence in the near-infrared by the similar comparative heights of their $Y$ and $J$ band peaks and a narrowing of the $J$ band flux peak \citep[Figure \ref{fig:Yspec} and ][]{cushing2011,kirkpatrick2012}. 
It is reasonable to note that the differences between the near-infrared spectra of late-T and Y~dwarfs show more subtle differences than seen across the LT transition. However, they also display significantly redder $J- W2$ colours and much fainter near-infrared magnitudes than late-T dwarfs. These differences suggest that the adoption of a new spectral type is appropriate. 

A review of the progress in characterising the Y dwarf population is provided in the chapter ``Y Dwarfs, the Challenge of Discovering the Coldest Substellar Population in the Solar Neighborhood'', by S. Leggett.
Given that the bulk of their emission escapes at wavelengths that are particularly challenging from the ground, detailed study of the Y~dwarfs will have to wait for the successful commissioning of JWST. 
However, initial characterisation based on parallaxes and the available limited near-infrared spectroscopy and multi-wavelength photometry suggest the Y~dwarfs have $T_{\rm eff}$ ranging from $\approx 500$~K down to $\approx 300$~K \citep[e.g.][]{leggett2017}.
Evolutionary models suggest that at typical thin disk ages of a few Gyr, these temperatures correspond to masses near, and below, the deuterium burning limit \citep[e.g.][]{baraffe2003}. 
As such, a significant proportion of the Y~dwarf population can also be classified as isolated planetary mass objects.

\subsection{The bottom of the IMF: Planetary mass objects}
Finding and studying the lowest mass brown dwarfs is compelling for a variety of reasons. 
As we've already discussed, determining the form of the IMF and the presence or otherwise of a low-mass cutoff is considered a key observable of the star formation process. 
This motivation has driven many searches for planetary mass brown dwarfs. 
In recent years, however, emphasis has shifted to look at how brown dwarfs can provide insights to understand the atmospheres of giant exoplanets \citep[e.g.][]{burgasser2011}. 
Although LT dwarfs span the same temperature range as giant exoplanets, they typically have larger masses, higher gravity and thus higher pressure photospheres. 
However, at ages of $\ltsimeq 500$Myr, planetary mass brown dwarfs occupy a wide range of spectral types on the LT sequence, and display similarly low surface gravity to that expected for giant exoplanets. 
To leverage this shared parameter space, a number of large-scale searches are ongoing to identify planetary mass brown dwarfs as members of young moving groups in the solar neighbourhood \citep[e.g.][]{aller2016,gagne2015}.

Planetary mass brown dwarfs have most often been identified initially by anomalously red $J - K_s$ colours \citep[e.g.][]{faherty2013,liu2013}. 
Spectroscopic signatures of low-gravity also highlight young, low-mass brown dwarfs \citep{allers2013}.
However, careful kinematic characterisation is then required to establish their membership of a young moving group to independently constrain their age \citep[e.g.][]{gagne2015}. 

Over 150 low-gravity L dwarfs are now known and their population property as a red and faint extension of the field L dwarf sequence is well established \citep{faherty2016}.
However, discoveries of T dwarf members of moving groups are few, and their observed spectral properties and colours do not obviously distinguish them from apparently older objects of similar type in the field \citep{naud2014, gagne2015b}. 
Large-scale kinematic searches are thus necessary to uncover the lowest mass contingent of young associations in the solar neighbourhood. 

\section{The near future}
\label{sec:future} 
The new generation of large-scale surveys have leveraged advances in optical design and imaging capabilities to achieve rapid coverage of the sky. 
This has opened the door to wide field synoptic surveys in the past few years such as VISTA VVV, Pan STARRS, and the under-construction Large Synoptic Survey Telescope \citep[LSST; ][]{lsst}. 
The LSST will survey the entire visible sky from  Cerro Pach\'on (Chile) every few nights. 
This observing strategy, aimed at discovering transients, will also provide proper motions and, more crucially, parallaxes for all the sources with measurable motions. 
This will open the door for a new unbiased method for searching for brown dwarfs through parallax selections. 
By selecting candidates based solely on their parallax and apparent luminosity, biases due to assumptions about colour and motion can be avoided. 
Such searches will likely find numerous brown dwarfs in the Solar neighbourhood that have been missed previously in regions such as the Galactic plane.

Also of note is the European Space Agency's Euclid mission \citep{euclid} which will provide deep optical and $YJH$ imaging along with slitless spectroscopy of the $1.1 - 2.0\mu$m region with $R \approx 250$ over large areas of sky.  Although targeted at extragalactic science, this mission will also provide the opportunity to study the brown dwarf population on sufficient scale to place them in a Galactic context.

\section{Cross-References}
\begin{itemize}
\item{First Discoveries of Brown Dwarfs and the Substellarity Tests}
\item{Y Dwarfs, the Challenge of Discovering the Coldest Substellar Population in the Solar Neighborhood}
\item{Spectral Properties of Brown Dwarfs and Unbound Planetary-Mass Objects}
\end{itemize}


\bibliographystyle{spbasicHBexo}  
\bibliography{refs} 

\end{document}